\documentclass[twocolumn]{aastex702} 

\usepackage{amsmath}
\usepackage{amssymb}
\usepackage{booktabs}
\usepackage{CJKutf8}

\DeclareFontShape{OT1}{cmr}{bx}{sc}{<->ssub*cmr/bx/n}{}

\graphicspath{{figures/}}

\newcommand{\athenak}{\textsc{AthenaK}}
\newcommand{\athenapp}{\textsc{Athena++}}
\newcommand{\dd}{\mathrm{d}}

\submitjournal{ApJS}
\shorttitle{Dynamical-Spacetime Radiation Transport}
\shortauthors{Zhu et al.}

\begin{document}
\begin{CJK*}{UTF8}{gbsn}

\title{Finite-Solid-Angle Boltzmann Radiation Transport on Dynamical Spacetimes in \athenak}

\author[orcid=0000-0001-9027-4184]{Hengrui Zhu ({\normalfont 朱恒锐})}
\altaffiliation{ \href{mailto:hengruizhu0330@gmail.com}{hengruizhu0330@gmail.com}}
\affiliation{Department of Physics, Princeton University, Princeton, NJ 08544, USA}
\affiliation{Princeton Gravity Initiative, Princeton University, Princeton, NJ 08544, USA}
\email{hengruizhu0330@gmail.com}

\author[orcid=0000-0001-6157-6722]{Alexander J. Dittmann}
\altaffiliation{NASA Einstein Fellow}
\affiliation{School of Natural Sciences, Institute for Advanced Study, 1 Einstein Drive, Princeton, NJ 08540, USA}
\email{dittmann@ias.edu}

\author[orcid=0000-0003-0232-0879]{Lizhong Zhang ({\normalfont 张力中})}
\affiliation{Center for Computational Astrophysics, Flatiron Institute, 162 Fifth Avenue, New York, NY 10010, USA}
\affiliation{School of Natural Sciences, Institute for Advanced Study, 1 Einstein Drive, Princeton, NJ 08540, USA}
\email{lizhong@ias.edu}

\author[orcid=0000-0001-5603-1832]{James M. Stone}
\affiliation{School of Natural Sciences, Institute for Advanced Study, 1 Einstein Drive, Princeton, NJ 08540, USA}
\email{jmstone@ias.edu}

\author[orcid=0000-0001-7941-801X]{Eduardo Mario Gutiérrez}
\affiliation{Department of Physics, The Pennsylvania State University, University Park, PA 16802, USA}
\affiliation{Institute for Gravitation and the Cosmos, The Pennsylvania State University, University Park, PA 16802, USA}
\email{exg5366@psu.edu}

\author[orcid=0000-0001-6982-1008]{David Radice}
\affiliation{Department of Physics, The Pennsylvania State University, University Park, PA 16802, USA}
\affiliation{Department of Astronomy and Astrophysics, The Pennsylvania State University, University Park, PA 16802, USA}
\affiliation{Institute for Gravitation and the Cosmos, The Pennsylvania State University, University Park, PA 16802, USA}
\email{david.radice@psu.edu}

\begin{abstract}
We extend the finite-solid-angle general relativistic radiation transport method of \citet{White2023} to time-dependent spacetimes represented in ADM form.
This Valencia-type solver retains the angular transport and local implicit matter coupling of the original HARM-type solver, but replaces its time-independent Kerr--Schild tetrad and conserved-variable normalization with an Eulerian formulation.
A Cholesky-gauge spatial tetrad makes the frame and its derivatives algebraic functions of the ADM variables, providing smooth, metric-compatible angular transport.
The resulting transport system couples directly to an evolving spacetime, agnostic to the gauge evolution equations.
Our test suite, including flat and curved beams, radiation--fluid coupling, and time-dependent geometries, establishes the accuracy and robustness of this approach.
We further apply the Valencia-type solver to a radiative circumbinary disk, illustrating its potential for multi-messenger studies of dynamical strongly gravitating systems.
\end{abstract}

\hfuzz=10pt
\section{Introduction}
\hfuzz=0.1pt

Radiation and neutrino transport shape the dynamics and observable signatures of many high-energy astrophysical systems.
In luminous black-hole accretion flows, radiation pressure and radiative cooling regulate the disk structure, launch outflows, and set the emergent spectra across nearly all accretion regimes, from X-ray binaries and active galactic nuclei to tidal disruption events \citep{DavisTchekhovskoy2020,Jiang2014SuperEdd,Jiang2019SuperEdd,Huang2023,Zhang2025RadI,Zhang2026RadII,Zhang:2026gfe,Dai2021TDEReview}.
Radiation hydrodynamics also shapes the turbulent envelopes and winds of massive stars, especially near opacity peaks \citep{Jiang2015,2018Natur.561..498J,2023ApJ...951L..42S,2025arXiv251014875M,2026ApJ...998L..10G}.
Neutrino transport is central to core-collapse supernovae: the neutrino-driven, turbulence-aided mechanism, descended from the classic neutrino-heating arguments of \citet{ColgateWhite1966}, is the leading explanation for core-collapse supernovae \citep{Janka2012,BurrowsVartanyan2021}, and neutrino--matter interactions set the electron fraction of the matter ejected by neutron-star mergers, thereby controlling $r$-process nucleosynthesis and the resulting kilonova emission \citep{Metzger2019,Foucart2023}.
In these applications, transport is often a leading modeling uncertainty.

The underlying problem is a Boltzmann transport equation for the photon or neutrino specific intensity.
The specific intensity depends on three spatial and three momentum dimensions, and even a frequency-integrated (gray) intensity still inhabits a computationally demanding five-dimensional phase space \citep[e.g.][]{Mihalas1984,Thorne1981,DavisGammie2020}.
Three broad families of algorithms make this problem tractable.
\emph{Moment methods} evolve a truncated hierarchy of angular moments \citep{Thorne1981,Shibata2011} closed by an assumed relation: flux-limited diffusion \citep{LevermorePomraning1981}, the M1/Minerbo closure \citep{Minerbo1978,Levermore1984}, or a variable Eddington tensor (VET), as in the hybrid Godunov framework of \citet{SekoraStone2010}, with the tensor obtained from a formal short-characteristics solution in \citet{JiangStoneDavis2012VET}.
This approach underlies most of the general relativistic radiation- and neutrino-transport codes in use today \citep{Sadowski2013,Sadowski2014,McKinney2014,Foucart2015,OConnor2015,Kuroda2016,Radice2022}.
\emph{Monte Carlo methods} sample the distribution function with computational packets and converge to the exact transport solution as the packet count grows \citep{Ryan2015,Foucart2018,FoucartMC2020,MillerRyanDolence2019,Foucart2021}; 
hybrid schemes close the moment hierarchy with transport information carried by deterministic samples or Monte Carlo packets, as in the method-of-characteristics moment closure of \citet{RyanDolence2020}, the Monte Carlo closure of \citet{Foucart2018Closure}, and the guided-moments formalism of \citet{Izquierdo2024}.
\emph{Deterministic angular methods} instead retain angular structure beyond low-order moments, either by evolving intensities along a fixed set of directions in discrete-ordinates (\(S_N\)) or finite-solid-angle schemes, or by expanding the angular dependence in a spherical-harmonic basis \citep{SumiyoshiYamada2012,Nagakura2014,Radice2013,BhattacharyyaRadice2023}.
Within the Athena family of codes this line began with the short-characteristics module of \citet{DavisStoneJiang2012}, continued with the time-dependent radiation-magnetohydrodynamics algorithms of \citet{Jiang2014} and \citet{Jiang2021}, and was extended to general relativity by \citet{White2023}, the immediate predecessor of the present work.

These methods do not lie on a single accuracy--cost hierarchy.
For example, the modeling error of an algebraic M1 closure does not vanish with spatial resolution.
M1 is inexpensive and performs well in the diffusion limit, but it merges crossing beams, can give an inaccurate Eddington tensor in semi-transparent regions between the diffusion and free-streaming limits, and can bias the neutrino density and pair-annihilation rate in merger polar funnels \citep{Sadowski2013,Foucart2018}.
VET methods avoid these algebraic-closure errors by computing the Eddington tensor from a formal solution of the radiative transfer equation over a set of discrete angles, at the cost of this additional angular transport solve \citep{JiangStoneDavis2012VET}.

Monte Carlo and deterministic Boltzmann methods instead have errors that can be reduced by increasing the packet count or angular resolution.
Monte Carlo methods resolve sharp angular features, but their sampling error decreases only as \(N^{-1/2}\) and can be expensive to reduce in optically thick regions \citep{Richers2017}.
Finite-solid-angle methods have neither an imposed moment closure nor sampling noise, and can resolve multiple beams and shadows by refining the angular grid.
Their main cost is storing and evolving many angular bins in each cell, a workload well suited to GPU-accelerated implementations such as \athenak\ \citep{Stone2024AthenaK,Zhang2025RadI}.

The \citet{White2023} solver, implemented in \athenapp\ and \athenak, evolves the frequency-integrated intensity on a geodesic angular mesh, uses upwind spatial and angular fluxes, accounts for gravitational light bending through tetrad rotation coefficients, and couples radiation to the fluid with a local implicit source update.
The performance-portable \athenak\ implementation has enabled radiation-(GR)MHD surveys of black-hole accretion over a wide range of accretion rates \citep{Zhang2025RadI,Zhang2026RadII,Zhang:2026gfe}.

The original GR radiation solver assumes a stationary analytic metric.
In the HARM-like formulation adopted by \citet{White2023}, following the stationary GRMHD scheme introduced by the HARM code \citet{Gammie2003}, the conserved intensity and its geometric source terms are written in terms of the coordinate metric and its derivatives, so the tetrad, angular advection coefficients, metric derivatives, and conserved-variable normalization can all be precomputed once for a fixed Cartesian Kerr--Schild (CKS) spacetime.
Extending the same construction to a time-dependent metric would require the coordinate metric's time derivatives in the source terms.
As \citet{Fields2024BNS} emphasize for the closely related GRMHD system, the analogous \(\partial_t g_{\alpha\beta}\) contribution to the energy source carries \(\partial_t\alpha\) and \(\partial_t\beta^i\), which cannot be eliminated without prior knowledge of the gauge conditions underlying the spacetime evolution; the HARM-like conserved variables therefore acquire an explicit gauge dependence on a dynamical spacetime.

Many relativistic systems of interest, including binary black holes, neutron-star mergers, and self-gravitating disks, involve dynamically evolving spacetimes.
The \athenak\ numerical relativity modules evolve ADM/Z4c metric fields on the same mesh \citep{Zhu2024NR,Fields2024BNS, Arnowitt:1962hi,Bernuzzi2010}, and, following the Valencia GRMHD formulation \citep{Banyuls1997}, we recast radiation transport on these backgrounds in a \(3+1\) decomposition. 
The resulting geometric source terms only require the instantaneous ADM fields and their spatial derivatives.

This paper describes the resulting dynamical-spacetime radiation solver.
Throughout, we refer to the original stationary-spacetime module---which evolves the HARM-like conserved variable \(k^0k_0I\)---as the \emph{HARM (radiation) solver}, and to the new dynamical-spacetime module built on ADM variables as the \emph{Valencia (radiation) solver}.
The Valencia solver retains the geodesic angular mesh, reconstruction, upwinding, and local matter-coupling machinery of \citet{White2023}; on a stationary background, the two formulations converge to the same continuum transport solution.
A Cholesky-gauge ADM tetrad fixes the otherwise arbitrary local spatial rotation by requiring the spatial co-triad to be the positive lower-triangular factor of the spatial metric \(\gamma_{ij}\).
As a result, the tetrad and the derivatives needed by the Hamiltonian angular-drift coefficients and matter-frame projections can be evaluated algebraically from ADM variables and their derivatives.
The remaining changes are the replacement of the stationary HARM-like normalized intensity by a Valencia-like densitized intensity, the ADM Hamiltonian angular drift, and the geometric energy source associated with lapse gradients and extrinsic curvature.
We focus on these differences rather than repeating the full derivation of the HARM solver.

The remainder of this paper is organized as follows.
In Section~\ref{sec:method}, we derive the Valencia finite-solid-angle radiation transport formulation for dynamical ADM spacetimes and describe its numerical implementation in \athenak.
Section~\ref{sec:tests} validates the method through a suite of transport, strong-field, radiation--matter coupling, and dynamical-spacetime benchmark problems.
Section~\ref{sec:application} presents a demonstration of the solver in a realistic astrophysical application: radiative accretion flow onto a binary black hole.
Section~\ref{sec:weak_scaling} discusses the scaling and performance of the implementation on GPU-accelerated supercomputers.
Finally, we summarize the main results and discuss future directions in Section~\ref{sec:summary}.

\section{Valencia radiation formulation}
\label{sec:method}

We formulate the dynamical-spacetime radiation solver using quantities measured by Eulerian observers on an ADM foliation.
We first review the equations for null geodesics in the context of the ADM formalism. Then we define phase-space variables and radiation moments. 
Following these preliminaries, we derive the conservative transport equation for photons in dynamical spacetimes and describe the finite-volume transport and local matter-coupling updates.
We conclude by relating the resulting Valencia formulation to the stationary HARM radiation solver of \citet{White2023}.

\subsection{Null geodesics in 3+1 form}
Following the standard ADM decomposition \citep{Gourgoulhon:2007ue}, the spacetime metric is written as
\begin{equation}
  \dd s^2
  =
  -\alpha^2\dd t^2
  +
  \gamma_{ij}
  \left(\dd x^i+\beta^i\dd t\right)
  \left(\dd x^j+\beta^j\dd t\right),
  \label{eq:adm_line_element}
\end{equation}
where $\gamma_{ij}$ is the three metric of the space-like hypersurface $\Sigma_t$ defined by constant coordinate time $t$, and $\alpha$ and $\beta^i$ are, respectively, the lapse function and the shift vector. The normal to the $\Sigma_t$ is
\begin{equation}
  n^\mu
  =
  \alpha^{-1}\left(1,-\beta^i\right), \quad
  n_\mu
  =
  \left(-\alpha,0,0,0\right).
\end{equation}
The Lie derivative of the three metric along $n^\mu$ is related to the extrinsic curvature by:
\begin{equation}
  K_{ij}
  =
  -\frac{1}{2}\mathcal{L}_n\gamma_{ij}.
\end{equation}
We decompose the four momentum of a photon as
\begin{equation}
  p^\mu
  =
  \epsilon\left(n^\mu+\ell^\mu\right),
  \quad
  n_\mu \ell^\mu
  =
  0,
  \quad
  \gamma_{ij} \ell^i \ell^j
  =
  1,
  \label{eq:photon_decomposition}
\end{equation}
where
\begin{equation}
  \epsilon
  =
  -p_\mu n^\mu
\end{equation}
is the photon energy measured by the Eulerian observer. Note that $\pmb{\ell}$ is purely spatial ($\ell^0 = 0$).

The velocity with which photons propagate in the Eulerian frame is
\begin{equation}
  v^i
 := 
  \frac{\dd x^i}{\dd t}
  =
  \frac{p^i}{p^0}
  =
  \alpha \ell^i-\beta^i.
  \label{eq:coordinate_light_speed}
\end{equation}

The energy of photons in the Eulerian frame evolves according to \citep{Vincent:2012kn}
\begin{equation}
  \frac{\dd \ln \epsilon}{\dd t} = \alpha K_{ij} \ell^i \ell^j - \ell^i \partial_i \alpha =: Q,
  \label{eq:photon.energy.evolution}
\end{equation}
where $t$ is the coordinate time (not an affine parameter along the photon trajectory). Also note that $Q$ is independent of $\epsilon$.

In the Eulerian frame, the photon position and momentum can be derived from the 3+1 form of the super Hamiltonian \citep{MTW}
\begin{equation}
  \mathcal{H} = \frac{1}{2} g^{\mu \nu} p_\mu p_\nu,
\end{equation}
which takes the form \citep{Schnittman:2003tm}
\begin{equation}
  H\left(x^i,p_i;t\right)
  =
  \alpha\epsilon-\beta^ip_i.
  \label{eq:null_hamiltonian}
\end{equation}
The corresponding equations of motion are
\begin{align}
  \dot{x}^i
  &=
  \frac{\partial H}{\partial p_i}
  =
  \alpha\frac{\gamma^{ij}p_j}{\epsilon}
  -
  \beta^i
  =
  \alpha \ell^i-\beta^i,
  \label{eq:hamilton_x}
  \\
  \frac{\dot{p}_i}{\epsilon}
  &=
  -\frac{1}{\epsilon}\frac{\partial H}{\partial x^i}
  \nonumber\\
  &=
  -\partial_i\alpha
  +
  \frac{p_j}{\epsilon}\partial_i\beta^j
  -
  \frac{\alpha}{2\epsilon^2}
  \left(\partial_i\gamma^{jk}\right)p_jp_k,
  \label{eq:hamilton_p}
\end{align}
where the dot denotes derivative with respect to coordinate time.

\subsection{Cholesky-gauge Eulerian tetrad}
Because the angular ordinates are defined in a local orthonormal frame, we must specify a tetrad to map each discrete direction to the coordinate basis.
As orthonormal spatial triads related by local rotations are physically equivalent but assign different coordinate directions to the same angular bins, this rotational freedom must be fixed to make the discrete angular grid and its derivatives unique and smooth.
We first fix the timelike tetrad leg to the Eulerian normal,
\begin{equation}
  e_{(\hat{0})}{}^\mu
  =
  n^\mu.
\end{equation}
The remaining spatial rotational freedom is fixed by a Cholesky factorization of the spatial metric:
\begin{equation}
  \gamma_{ij}
  =
  \sum_{\hat{a}=1}^{3}L_{i\hat{a}}L_{j\hat{a}},
  \qquad
  L
  =
  \begin{pmatrix}
    L_{11} & 0      & 0 \\
    L_{21} & L_{22} & 0 \\
    L_{31} & L_{32} & L_{33}
  \end{pmatrix},
  \label{eq:cholesky_factorization}
\end{equation}
with positive diagonal entries.
Explicitly,
\begin{align}
  L_{11}
  &=
  \sqrt{\gamma_{11}},
  ~~
  L_{21}
  =
  \frac{\gamma_{12}}{L_{11}},
  ~~
  L_{31}
  =
  \frac{\gamma_{13}}{L_{11}},
  \nonumber\\
  L_{22}
  &=
  \sqrt{\gamma_{22}-L_{21}^2},
  \nonumber\\
  L_{32}
  &=
  \frac{\gamma_{23}-L_{31}L_{21}}{L_{22}},
  ~~
  L_{33}
  =
  \sqrt{\gamma_{33}-L_{31}^2-L_{32}^2}.
  \label{eq:cholesky_entries}
\end{align}

The spatial co-triad and dual triad can then be constructed as
\begin{equation}
  e^{(\hat{a})}{}_i
  =
  L_{i\hat{a}},
  \qquad
  e_{(\hat{a})}{}^i
  =
  (L^{-1})_{\hat{a}i}.
  \label{eq:triad_definitions}
\end{equation}
Equivalently, when the coordinate index labels the matrix row, the matrix whose columns are the dual-triad vectors is \(L^{-T}\).
These fields satisfy
\begin{equation}
  e_{(\hat{a})}{}^ie^{(\hat{b})}{}_i
  =
  \delta_{\hat{a}}{}^{\hat{b}},
  \qquad
  \gamma_{ij}
  e_{(\hat{a})}{}^i
  e_{(\hat{b})}{}^j
  =
  \delta_{\hat{a}\hat{b}},
  \label{eq:triad_orthonormality}
\end{equation}
where the second equation follows from substituting Equation~\eqref{eq:triad_definitions} into Equation~\eqref{eq:cholesky_factorization}.

An angular-grid direction in the orthonormal tetrad frame, $\ell^{(\hat{a})}$, is then mapped to the coordinate basis according to
\begin{equation}
  \ell^i
  =
  e_{(\hat{a})}{}^i\ell^{(\hat{a})},
  \qquad
  \frac{p_i}{\epsilon}
  =
  \ell_i
  =
  e^{(\hat{a})}{}_i\ell_{(\hat{a})}.
  \label{eq:direction_triad_map}
\end{equation}
Importantly, the components of $\ell^{(\hat{a})}$ represent the direction of propagation of the photon in an orthonormal frame comoving with the Eulerian observer. In particular, we can write
\begin{equation}
  \ell^{(\hat{a})} = \left( 
    \cos \varphi \sin \theta,\,
    \sin \varphi \sin \theta,\,
    \cos \theta
  \right)
\end{equation}
for angles $0 \leq \theta \leq \pi$, $0 \leq \varphi < 2\pi$.

For any derivative \(\partial_\lambda\), differentiating Equation~\eqref{eq:cholesky_factorization} yields 
\begin{equation}
  \partial_\lambda\gamma_{ij}
  =
  \sum_{\hat{a}=1}^{3}
  \left[
    (\partial_\lambda L_{i\hat{a}})L_{j\hat{a}}
    +
    L_{i\hat{a}}(\partial_\lambda L_{j\hat{a}})
  \right],
  \label{eq:cholesky_derivative}
\end{equation}
which uniquely determines the derivatives of the lower-triangular matrix \(\partial_\lambda L\), and thereby those of the tetrads.
We compute the spatial derivatives by finite-differencing the numerical spatial metric directly, while the time derivative follows from the ADM evolution equation
\begin{equation}
  \partial_t\gamma_{ij}
  =
  -2\alpha K_{ij}
  +
  D_i\beta_j
  +
  D_j\beta_i.
  \label{eq:adm_metric_time_derivative}
\end{equation}
Here, \(D_i\) denotes the spatial covariant derivative associated with the spatial metric.
The tetrad time derivative therefore requires neither \(\partial_t\alpha\) nor \(\partial_t\beta^i\).

\subsection{Radiation intensity and moments}
Up to a factor of the speed of light $c$, we define the radiation intensity as the energy density of photons $\dd E$ in a volume $\dd V$, traveling in the direction subtended by the solid angle $\dd \Omega$, per unit frequency $\dd \nu$:
\begin{equation}
   I_\nu = \frac{c\, \dd E}{\dd V\, \dd \Omega\, \dd \nu}.
\end{equation}
In the tetrad frame, $\dd \Omega$ takes the usual form $\dd \Omega = \sin \theta\, \dd \theta\, \dd \varphi$, $h \nu = \epsilon$, and the spatial volume form is $\dd V = \sqrt{\gamma}\, \dd^3 x$, where $\gamma$ is the determinant of $\gamma_{ij}$. 

The gray (energy-integrated) intensity is then
\begin{equation}
  I\left(t,x^i,\ell^{(\hat{a})}\right)
  =
  \int_0^\infty I_\nu\,\dd\nu,
  \label{eq:valencia_primitive_intensity}
\end{equation}
from which we compute the Eulerian angular moments
\begin{equation}
  E
  =
  \int I\,\dd\Omega,
  ~~
  F^i
  =
  \int I\ell^i\,\dd\Omega,
  ~~
  P^{ij}
  =
  \int I\ell^i\ell^j\,\dd\Omega,
  \label{eq:eulerian_moments}
\end{equation}
so that the radiation stress-energy tensor can be written as
\begin{equation}
  R^{\mu\nu}
  =
  En^\mu n^\nu
  +
  F^\mu n^\nu
  +
  n^\mu F^\nu
  +
  P^{\mu\nu},
  \label{eq:radiation_stress_energy}
\end{equation}
where
\begin{equation}
  F^\mu n_\mu=0,
  \qquad
  P^{\mu\nu}n_\nu=0.
\end{equation}
The finite-volume variable evolved by the Valencia radiation solver is the densitized Eulerian intensity:
\begin{equation}
  U
  =
  \sqrt{\gamma}\,I.
  \label{eq:valencia_cons_intensity}
\end{equation}
An evolution equation for the latter is obtained from the conservative transport equation, as discussed in the next section.

\subsection{Conservative gray transport equation}
In the absence of collisions, the invariant distribution function $f_\nu = I_\nu/\nu^3$ is conserved along null geodesics. 
To derive an evolution equation for the gray intensity, we account for the changing photon frequency by considering the invariant frequency measure $\dd\ln\nu$. 
The resulting quantity $f_\nu\dd\ln\nu=\frac{I_\nu}{\nu^4}\dd\nu$ is therefore conserved along collisionless photon trajectories:
\begin{equation}
  \frac{\dd}{\dd t} \left[ \frac{I_\nu \dd \nu}{\nu^4} \right] =
  \frac{1}{\nu^4} \left[ \frac{\dd I_\nu}{\dd t} - 4 I_\nu \frac{\dd \ln \nu}{\dd t} \right] \dd \nu = 0.
\end{equation}
Integrating over frequencies and using \eqref{eq:photon.energy.evolution}, we obtain the non-conservative form of the transport equation:
\begin{equation}
  \frac{\dd I}{\dd t} = \left[ \frac{\partial}{\partial t} + v^i \frac{\partial}{\partial x^i} + \dot{\ell}^{(\hat{a})} \frac{\partial}{\partial \ell^{(\hat{a})}} \right] I = 4 I Q + \partial_t I |_{\rm coll}.
  \label{eq:gray_nonconservative}
\end{equation}
In the last equation, we have also included the collisional term $\partial_t I |_{\rm coll}$, which describes interaction between photons and matter.

To obtain the transport equation in conservation form, we follow an approach similar to that of \citet{DavisGammie2020}. Remembering that the photon number is \citep[e.g.][]{Mihalas1984}
\begin{equation}
   \dd N = 2 f_\nu\, \sqrt{\gamma} \dd^3 x\, \epsilon^2 \dd \epsilon\, \dd \Omega,
\end{equation}
conservation of photon number (in absence of interaction with the medium) implies that
\begin{equation}
\begin{split}
  &\frac{\partial (\sqrt{\gamma} f_\nu)}{\partial t} + \frac{\partial (\sqrt{\gamma} f_\nu \dot{x}^i)}{\partial x^i}
  + \\ & \qquad \frac{1}{\epsilon^2} \frac{\partial (\sqrt{\gamma} f_\nu \dot{\epsilon} \epsilon^2)}{\partial \epsilon} +
  \nabla_\Omega \cdot (\sqrt{\gamma} f_\nu \dot{\boldsymbol{\ell}}) = 0,
\end{split}
\end{equation}
where an overdot denotes derivative with respect to time along the photon trajectories, e.g., $\dot{x}^i = v^i$. Together with the invariance of $f_\nu$ along the particle trajectories,
\begin{equation}
  \frac{\dd f_\nu}{\dd t} = 0,
\end{equation}
this yields Liouville's theorem:
\begin{equation}
  \partial_t\ln\sqrt{\gamma}
  +
  \frac{1}{\sqrt{\gamma}}
  \partial_i\left(\sqrt{\gamma}\,v^i\right)
  +
  \nabla_\Omega\cdot\dot{\boldsymbol{\ell}}
  =
  -3Q.
  \label{eq:phase_space_divergence}
\end{equation}
Combining Equations~\eqref{eq:valencia_cons_intensity}, \eqref{eq:gray_nonconservative}, and \eqref{eq:phase_space_divergence} yields the conservative equation advanced by the Valencia solver:
\begin{align}
  \partial_t U
  &+
  \partial_i
  \left[
    \left(\alpha \ell^i-\beta^i\right) U
  \right]
  +
  \nabla_\Omega\cdot
  \left(U \dot{\boldsymbol{\ell}} \right)
  \nonumber\\
  &=
  Q U
  +
  S^{\rm coll},
  \label{eq:adm_angle_transport}
\end{align}
here $Q$ encodes the gravitational and Doppler redshift/blueshift, and
\begin{equation}
  S^{\rm coll}
  :=
  \sqrt{\gamma}
  \left.\partial_t I \right|_{\rm coll}
  \label{eq:densitized_collision_source}
\end{equation}
encodes the coupling with matter and is positive when matter transfers energy into the radiation field.

The zeroth and first angular moments of Equation~\eqref{eq:adm_angle_transport} recover the Valencia radiation energy and momentum equations:
\begin{align}
  \partial_t\left(\sqrt{\gamma}E\right)
  &+
  \partial_i
  \left[
    \sqrt{\gamma}
    \left(\alpha F^i-\beta^iE\right)
  \right]
  \nonumber\\
  &=
  \sqrt{\gamma}
  \left(
    \alpha P^{ij}K_{ij}
    -
    F^i\partial_i\alpha
    +
    \mathcal{S}_E
  \right),
  \label{eq:valencia_rad_energy}
  \\
  \partial_t\left(\sqrt{\gamma}F_i\right)
  &+
  \partial_j
  \left[
    \sqrt{\gamma}
    \left(\alpha P_i{}^j-\beta^jF_i\right)
  \right]
  \nonumber\\
  &=
  \sqrt{\gamma}
  \left(
    -E\partial_i\alpha
    +
    F_j\partial_i\beta^j
    +
    \frac{\alpha}{2}
    P^{jk}\partial_i\gamma_{jk}
    +
    \mathcal{S}_i
  \right),
  \label{eq:valencia_rad_momentum}
\end{align}
where the coordinate-time collision source moments are defined by
\begin{equation}
  \sqrt{\gamma}\,\mathcal{S}_E
  =
  \int S^{\rm coll}\,\dd\Omega,
  \qquad
  \sqrt{\gamma}\,\mathcal{S}_i
  =
  \int \ell_iS^{\rm coll}\,\dd\Omega.
  \label{eq:collision_source_moments}
\end{equation}
The quantities \(\mathcal{S}_E\) and \(\mathcal{S}_i\) are defined here as the source terms appearing in the coordinate-time moment equations; they should not be confused directly with coordinate components of a covariant four-force without the corresponding lapse and projection factors.

The angular derivative terms include both the momentum source and the motion of the local tetrad.
Define the derivative along the coordinate-space characteristic,
\begin{equation}
  \mathcal{D}_t
  :=
  \partial_t+v^j\partial_j,
\end{equation}
and the projector tangent to the unit sphere,
\begin{equation}
  \mathcal{P}^{\hat{a}}{}_{\hat{b}}
  =
  \delta^{\hat{a}}{}_{\hat{b}}
  -
  \ell^{(\hat{a})}\ell_{(\hat{b})}.
\end{equation}
Differentiating
\begin{equation}
  p_i
  =
  \epsilon\, e^{(\hat{a})}{}_i\ell_{(\hat{a})}
\end{equation}
and projecting orthogonally to \(\ell^{(\hat{a})}\) gives
\begin{equation}
  \dot{\ell}^{(\hat{a})}
  =
  \mathcal{P}^{\hat{a}}{}_{\hat{b}}
  e_{(\hat{b})}{}^i
  \left[
    \frac{\dot{p}_i}{\epsilon}
    -
    \left(
      \mathcal{D}_te^{(\hat{c})}{}_i
    \right)
    \ell_{(\hat{c})}
  \right].
  \label{eq:angular_characteristic}
\end{equation}
The projector removes the component associated with the change in momentum magnitude and guarantees
\begin{equation}
  \ell_{(\hat{a})}\dot{\ell}^{(\hat{a})}
  =
  0.
\end{equation}
Thus \(\dot{\boldsymbol{\ell}}\) is tangent to \(S^2\).
In the implementation, the co-triad is differentiated in time and space, Equation~\eqref{eq:hamilton_p} supplies the covariant momentum force, and the resulting tangent vector is projected onto each angular-cell edge to construct the upwind angular flux.

\subsection{Finite volume discretization}
Following \citet{White2023}, we discretize Equation~\eqref{eq:adm_angle_transport} on a spherical grid. For angular cell $\Omega_n$ with weight $w_n=|\Omega_n|$, the averaged conserved intensity is
\begin{equation}
  U_n = \frac{1}{w_n}\int_{\Omega_n} I \sqrt{\gamma} \,\dd\Omega.
\end{equation}
The corresponding densitized discrete moments can be computed as
\begin{align}
  \mathcal{E}
  &:=
  \sum_nw_nU_n
  =
  \sqrt{\gamma}\,E,
  \nonumber\\
  \mathcal{F}_i
  &:=
  \sum_nw_n\ell_{i,n}U_n
  =
  \sqrt{\gamma}\,F_i,
  \nonumber\\
  \mathcal{P}_i{}^j
  &:=
  \sum_nw_n\ell_{i,n}\ell_n^jU_n
  =
  \sqrt{\gamma}\,P_i{}^j.
  \label{eq:densitized_discrete_moments}
\end{align}

In the implementation, the ADM fields, inverse spatial metric, determinant, tetrad, and co-triad are stored at cell centers before radiation transport.
For an evolving Z4c spacetime or a prescribed time-dependent ADM spacetime, these cell-centered quantities are refreshed at each Runge--Kutta stage.
At each spatial face, the face-centered spatial metric is factorized independently, and the geometric quantities stored there are the transport coefficients
\begin{equation}
  \left(
    -\beta^d,\,
    \alpha e_{(\hat{1})}{}^d,\,
    \alpha e_{(\hat{2})}{}^d,\,
    \alpha e_{(\hat{3})}{}^d
  \right)
\end{equation}
together with \(\sqrt{\gamma}\).
Constructing and storing these cell- and face-centered geometric quantities costs \(O(N_{\rm cell})\) per metric refresh.
Evaluating the direction-dependent spatial and angular transport coefficients subsequently costs \(O(N_{\rm cell}N_{\rm ang})\).
The lapse, shift, and six independent components of \(\gamma_{ij}\) at a face are reconstructed from neighboring cell-centered values.
The face metric is then factorized directly to form the transport coefficients and~\(\sqrt{\gamma}\).

\subsection{Finite-volume update}

The primitive reconstructed in each angular bin is \(I_n=U_n/\sqrt{\gamma}\).
For a spatial face normal to \(x^d\), the numerical flux is
\begin{equation}
  \mathcal{F}_{n,f}^{\,d}
  =
  \sqrt{\gamma_f}\,
  v_{n,f}^{\,d}
  I_{n,f}^{\rm up},
  ~~
  v_{n,f}^{\,d}
  =
  -\beta_f^d
  +
  \alpha_f
  e_{(\hat{a}),f}{}^d
  \ell_n^{(\hat{a})},
  \label{eq:spatial_numerical_flux}
\end{equation}
where the sign of \(v_{n,f}^{\,d}\) selects the upwind state.

On the geodesic angular mesh, let \(\widehat{\boldsymbol{m}}_{nq}\) be the outward unit tangent-space conormal to edge \(q\) of angular cell \(n\), and let \(\Delta\sigma_{nq}\) be the edge length.
The integrated flux through the edge is
\begin{equation}
  \mathcal{A}_{nq}
  =
  \Delta\sigma_{nq}
  \left(
    \dot{\boldsymbol{\ell}}_{nq}
    \cdot
    \widehat{\boldsymbol{m}}_{nq}
  \right)
  U_{nq}^{\rm up},
  \label{eq:angular_numerical_flux}
\end{equation}
and the angular divergence is approximated by
\begin{equation}
  \left[
    \nabla_\Omega\cdot
    \left(U\dot{\boldsymbol{\ell}}\right)
  \right]_n
  =
  \frac{1}{\Delta\Omega_n}
  \sum_q\mathcal{A}_{nq},
  \label{eq:angular_flux_divergence}
\end{equation}
where \(\Delta\Omega_n\) is the solid angle of angular cell \(n\).

Spatial and angular transport are advanced with an explicit Runge--Kutta integrator.
Within each stage, the geometric source term \(Q_n\) is frozen at the stage geometry.
The implementation adds the exponential increment
\begin{equation}
  \Delta U_n^{\rm geom}
  =
  U_n^{\rm stage}
  \left[
    \exp\left(\Delta t_{\rm stage}Q_n\right)-1
  \right]
  \label{eq:exponential_metric_source}
\end{equation}
to the Runge--Kutta transport update.
For the isolated linear equation \(\partial_tU_n=Q_nU_n\), this is the exact frozen-\(Q_n\) increment over \(\Delta t_{\rm stage}\).

The local radiation--matter solve is operator-split from the explicit transport update.
Consequently, although the transport discretization may be second-order or higher, the fully coupled method is generally asymptotically first-order in time.

\subsection{Radiation--matter coupling}

The local source solve is performed in the fluid-comoving frame.
If \(u^{(\hat{\alpha})}\) is the fluid four-velocity in the Eulerian tetrad, the Doppler factor of ordinate \(n\) is
\begin{equation}
  \mathcal{D}_n
  =
  u^{(\hat{0})}
  -
  u^{(\hat{a})}\ell_{n(\hat{a})}
  =
  -\frac{u_\mu p^\mu}{\epsilon}.
  \label{eq:doppler_factor}
\end{equation}
Frequency-integrated intensity and solid angle transform as
\begin{equation}
  I_{{\rm cm},n}
  =
  \mathcal{D}_n^4I_n,
  \qquad
  \dd\Omega_{{\rm cm},n}
  =
  \mathcal{D}_n^{-2}\dd\Omega_n.
  \label{eq:intensity_lorentz_transform}
\end{equation}
Because \(p^0=\epsilon/\alpha\), a coordinate-time interval corresponds to the comoving photon path length
\begin{equation}
  \Delta\ell_{{\rm cm},n}
  =
  -u_\mu p^\mu\frac{\Delta t}{p^0}
  =
  \alpha\mathcal{D}_n\Delta t.
  \label{eq:comoving_optical_path}
\end{equation}

For absorption, elastic isotropic scattering, and distinct Planck- and Rosseland-mean opacities, we define the non-Compton gray collision operator as
\begin{align}
  \mathcal{C}^{(0)}_{{\rm cm},n}
  ={}&
  \sigma_a
  \left(
    B-I_{{\rm cm},n}
  \right)
  +
  \sigma_s
  \left(
    J_{\rm cm}-I_{{\rm cm},n}
  \right)
  \nonumber\\
  &+
  \sigma_p
  \left(
    B-J_{\rm cm}
  \right),
  \label{eq:comoving_collision_operator}
\end{align}
where
\begin{equation}
  B
  =
  \frac{a_{\rm rad}T_{\rm gas}^4}{4\pi},
  \qquad
  J_{\rm cm}
  =
  \frac{1}{4\pi}
  \int I_{\rm cm}\,\dd\Omega_{\rm cm}.
\end{equation}
Here \(\sigma_a\) and \(\sigma_s\) are the comoving absorption and scattering coefficients, while \(\sigma_p\) is the Planck-minus-Rosseland correction.

In the discrete source solve, aberration changes the angular weights to \(w_n\mathcal{D}_n^{-2}\), so we use the normalized quadrature
\begin{equation}
  J_{\rm cm}
  \simeq
  \frac{\sum_nw_n\mathcal{D}_n^{-2}I_{{\rm cm},n}}
       {\sum_nw_n\mathcal{D}_n^{-2}}.
  \label{eq:discrete_comoving_mean}
\end{equation}

The first local implicit update solves
\begin{equation}
  I_{{\rm cm},n}^{*}
  =
  I_{{\rm cm},n}^{-}
  +
  \alpha\Delta t\,\mathcal{D}_n
  \mathcal{C}^{(0)}_{{\rm cm},n}
  \left(
    I_{\rm cm}^{*},
    T_{\rm gas}^{*}
  \right).
  \label{eq:implicit_comoving_update}
\end{equation}
All angular bins and the gas temperature are coupled through \(J_{\rm cm}\) and energy conservation.
Temperature-dependent opacities are reevaluated until the relative temperature change satisfies the requested tolerance.

When enabled, Compton energy exchange is applied in a subsequent, separate local implicit update, so the non-Compton and Compton source operators are Lie-split.
This gray prescription follows the frequency-integrated thermal Compton approximation of \citet{Jiang2014Corona}: it assumes Thomson scattering by nonrelativistic thermal electrons and represents the radiation spectrum by the single temperature \(T_{\rm rad}\).
The comoving gray Compton source added isotropically to each ordinate is
\begin{equation}
\begin{split}
  \mathcal{C}^{\rm C}_{{\rm cm},n}
  & =
  4\sigma_sJ_{\rm cm}
  \frac{T_{\rm gas}-T_{\rm rad}}{T_e},\\
  T_{\rm rad}
  & =
  \left(\frac{4\pi J_{\rm cm}}{a_{\rm rad}}\right)^{1/4},
  \qquad
  T_e
  =
  \frac{m_ec^2}{k_{\rm B}}.
\end{split}
  \label{eq:compton_source}
\end{equation}
The implicit update solves the resulting quartic for the new radiation temperature, obtains the gas temperature from cell-local gas--radiation energy conservation, and applies the corresponding intensity increment to every angular bin.

\begin{table*}[t]
\centering
\caption{Comparison of the HARM-like and Valencia radiation formulations.}
\label{tab:harm_valencia}
\begin{tabular}{lll}
\toprule
 & \textbf{HARM-like} & \textbf{Valencia} \\
\midrule
Spacetime
  & stationary \(g_{\mu\nu}\)
  & dynamical ADM variables (\(\alpha,~\beta^i,~\gamma_{ij}\))\\

Photon energy
  & Killing energy: \(-k_0\)
  & Eulerian energy: \(\epsilon=-p_\mu n^\mu\) \\

Conservative variable
  & \(k^0 k_0 I_n\)
  & \(\sqrt{\gamma}\,I_n\) \\

Primitive variable
  & \(k_0 I_n\)
  & \(I_n\) \\

Spatial velocity
  & \(k^i/k^0\)
  & \(\alpha\ell^i-\beta^i\) \\

Tetrad
  & stationary triangular
  & time-dependent Cholesky factorization \\

Geometry
  & precomputed
  & updated each RK stage \\

Matter coupling
  & comoving-frame implicit
  & comoving-frame implicit \\
\bottomrule
\end{tabular}
\end{table*}

Feedback to the Valencia fluid variables is computed from the actual change of the discrete radiation moments rather than from a separately approximated four-force.
Superscripts \(-\) and \(+\) denote states immediately before and after the local source update, respectively.
For the matter stress--energy tensor \(T_{\rm m}^{\mu\nu}\), define \(E_{\rm m}=n_\mu n_\nu T_{\rm m}^{\mu\nu}\), \(S_i=-\gamma_{i\mu}n_\nu T_{\rm m}^{\mu\nu}\), and \(D=\rho W\), where \(W=-n_\mu u^\mu\).
The Valencia energy variable used here is \(\tau=E_{\rm m}-D\).
For each local source update, defining the pre- and post-update moments by Equation~\eqref{eq:densitized_discrete_moments}, we apply
\begin{align}
  \left(\sqrt{\gamma}\tau\right)^{+}
  &=
  \left(\sqrt{\gamma}\tau\right)^{-}
  +
  \mathcal{E}^{-}
  -
  \mathcal{E}^{+},
  \label{eq:discrete_energy_exchange}
  \\
  \left(\sqrt{\gamma}S_i\right)^{+}
  &=
  \left(\sqrt{\gamma}S_i\right)^{-}
  +
  \mathcal{F}_i^{-}
  -
  \mathcal{F}_i^{+}.
  \label{eq:discrete_momentum_exchange}
\end{align}
The baryon density is unchanged, so the radiation-energy increment is also the appropriate increment to the Valencia energy variable \(\sqrt{\gamma}\tau\).
For a successful local source update with fluid feedback enabled, Equations~\eqref{eq:discrete_energy_exchange} and \eqref{eq:discrete_momentum_exchange} conserve the discrete cell-local sum of matter and radiation Eulerian energy and all three covariant momenta to roundoff.

\subsection{Relation to the stationary finite-solid-angle method}
\label{sec:stationary}

The stationary and dynamical-spacetime solvers use the same geodesic angular mesh, spatial and angular upwinding, and local implicit radiation--matter coupling.
Their essential difference is the finite-volume variable used for each angular bin and, consequently, the geometric terms required to transport it.

Let
\begin{equation}
  k^\mu
  =
  e_{(\hat{\alpha})}{}^\mu\ell^{(\hat{\alpha})},
  ~~
  \ell^{(\hat{\alpha})}
  =
  \left(1,\ell^{(\hat{a})}\right),
  ~~
  \delta_{\hat{a}\hat{b}}
  \ell^{(\hat{a})}\ell^{(\hat{b})}
  =
  1,
\end{equation}
denote a future-directed null direction with unit frequency in the local orthonormal frame.
On a stationary Cartesian Kerr--Schild background, the HARM radiation solver evolves
\begin{equation}
  U_n^{\rm H}
  =
  k^0k_0I_n.
  \label{eq:harm_cons_intensity}
\end{equation}
Spatial reconstruction is applied to
\begin{equation}
  k_0I_n
  =
  \frac{U_n^{\rm H}}{k^0},
\end{equation}
while recovery of \(I_n\) and construction of radiation moments require division by \(k^0k_0\).
This choice is natural on a stationary metric because \(-k_0\) is the photon energy associated with the Killing vector \((\partial_t)^\mu\), and the metric, tetrad, angular-advection coefficients, and conserved-variable normalization can be precomputed.

A direct extension of Equation~\eqref{eq:harm_cons_intensity} to an evolving spacetime introduces \(\partial_tg_{\mu\nu}\), including \(\partial_t\alpha\) and \(\partial_t\beta^i\), into the transport source.
Eliminating these terms would require explicit knowledge of the lapse and shift evolution equations rather than only the instantaneous ADM geometry.
By contrast, the Valencia variable \(U_n=\sqrt{\gamma}I_n\), defined in Equation~\eqref{eq:valencia_cons_intensity}, leads to geometric terms written entirely in terms of \(\alpha\), \(\beta^i\), \(\gamma_{ij}\), \(K_{ij}\), and their spatial derivatives.
No separate evolution equations for \(\alpha\) or \(\beta^i\) are required.

On a stationary background, the HARM and Valencia formulations evolve different finite-volume normalizations of the same continuum intensity.
In Cartesian Kerr--Schild coordinates, both tetrads use the Eulerian normal as the timelike leg, but their triangular spatial gauges bend different in-plane legs.
The HARM tetrad fixes \(e_{(\hat{2})}{}^i\) along the coordinate \(y\) direction, whereas the Valencia tetrad fixes \(e_{(\hat{1})}{}^i\) along the coordinate \(x\) direction.
After accounting for this local rotation, the two solvers converge to the same radiation field. 
A side-by-side comparison between the two schemes is summarized in Table~\ref{tab:harm_valencia}.

\begin{figure}[t]
\centering
\includegraphics[width=\linewidth]{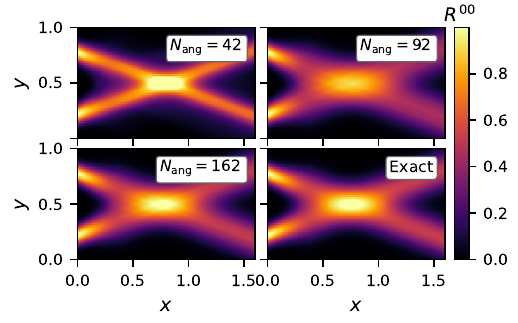}
\caption{Crossing-beams coordinate-frame radiation component \(R^{00}\) at three angular resolutions and in the exact free-streaming solution.
The numerical solution approaches the exact broadened profile as \(N_{\rm ang}\) increases, while the beams cross without merging.}
\label{fig:crossing_maps}
\end{figure}

\begin{figure}[htbp]
\centering
\includegraphics[width=\linewidth]{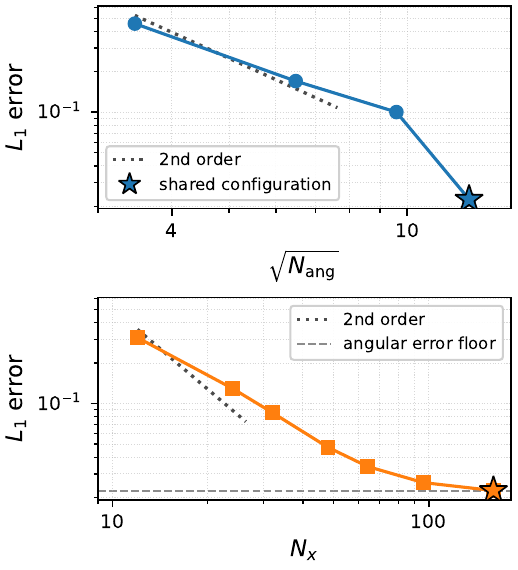}
\caption{Relative \(L_1\) error of the crossing-beams solution under angular refinement at fixed \(160\times100\) spatial resolution (top) and spatial refinement at fixed \(N_{\rm ang}=162\) (bottom).
The star is the shared finest configuration, and the dashed line marks the angular-error floor of the spatial sweep.}
\label{fig:crossing_conv}
\end{figure}

\section{Numerical tests}
\label{sec:tests}

\subsection{Crossing beams}

The crossing-beams problem tests multidirectional free streaming and separates angular from spatial diffusion through comparison with a simple analytic solution.
We use a quasi-two-dimensional domain \((x,y)\in[0,1.6]\times[0,1]\), with one periodic cell in the third direction.
Two beams with unit peak intensity and transverse Gaussian width \(\sigma=0.055\) originate on axes through \((-0.2,0.15)\) and \((-0.2,0.85)\), both directed toward \((0.75,0.5)\).
Their intensities fill the domain initially and are continuously injected through the \(x=0\) boundary; the remaining in-plane boundaries are set to outflow.
Each beam has a maximum-entropy angular distribution with flux factor \(0.95\).

All runs use piecewise-linear reconstruction, RK2 time integration, and a CFL number of \(0.3\).
For the angular-resolution study, we fix the spatial grid at \(160\times100\) cells and use \(N_{\rm ang}=12,42,92,162\).
For the spatial-resolution study, we fix \(N_{\rm ang}=162\) and use \(N_x=12,24,32,48,64,96,160\), with \(N_y=5N_x/8\).
Figure~\ref{fig:crossing_maps} compares the exact free-streaming solution with the numerical results for \(N_{\rm ang}=42\), \(92\), and \(162\).
The two beams pass through one another without merging, and the numerical profiles approach the exact solution as the angular resolution increases.

Figure~\ref{fig:crossing_conv} shows the relative \(L_1\) error in the coordinate-frame radiation component \(R^{00}\).
The top panel varies the angular resolution at fixed spatial resolution, while the bottom panel varies the spatial resolution at fixed \(N_{\rm ang}=162\).
We plot the angular sweep against \(\sqrt{N_{\rm ang}}\), the effective angular resolution on the x-y plane, so that a second-order error scales with slope \(-2\).
The two sweeps share the finest configuration, \((N_{\rm ang}=162,\,160\times100)\), marked by a star in both panels.
The angular error decreases at approximately second order and reaches \(2.3\%\) at the highest resolution.
The spatial error initially converges at a rate between first and second order, but levels off near \(2\%\) once it reaches the discretization error of the fixed angular grid.

\begin{figure*}[t]
\centering
\includegraphics[width=\textwidth]{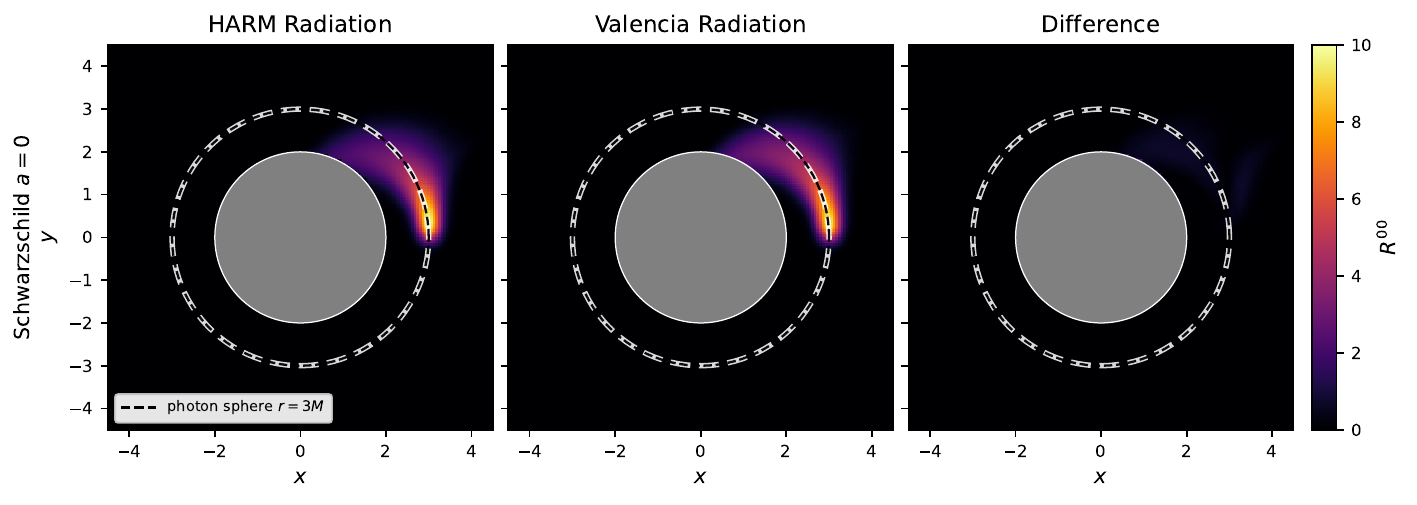}
\caption{Photon-ring beam on a stationary Schwarzschild background: the coordinate-frame radiation component \(R^{00}\) from the HARM solver (left), the Valencia solver (middle), and their absolute pointwise difference (right).
The beam is injected tangent to the photon sphere at \(r=3M\) (dashed circle).}
\label{fig:kerr_beam}
\end{figure*}

\subsection{Photon-ring beam on a Schwarzschild background}

The crossing-beam problem tests transport in flat spacetime. 
We now add strong curvature by following a beam near the Schwarzschild photon sphere.
Because the beam remains close to the unstable circular photon orbit, any small errors in the tetrad or angular geodesic advection accumulate and become readily visible.
As discussed in Section~\ref{sec:stationary}, the HARM and Valencia tetrads have the same Eulerian timelike leg in Cartesian Kerr--Schild coordinates but use different spatial orientations.
Their angular-bin intensities therefore cannot be compared directly, so we instead compare again the common coordinate-frame component \(R^{00}\).

We use a Schwarzschild background (\(a=0\)) and \(N_{\rm ang}=162\).
The quasi-two-dimensional domain spans \([-4.5M,4.5M]^2\) with \(144^2\) cells and one periodic cell in the third direction.
A continuous beam with amplitude \(8\) and transverse Gaussian width \(0.18M\) is injected tangentially to the photon sphere at \(r=3M\).
Both solvers use piecewise-linear reconstruction, RK2 time integration, and a CFL number of \(0.25\), and are compared at \(t=6M\).
Figure~\ref{fig:kerr_beam} shows the coordinate-frame component \(R^{00}\) from the two solvers and their pointwise difference.
At the present resolution, the pointwise difference is less than 5\%, and it converges away with resolution.

\subsection{Gas-radiation equilibration}

Following the free-streaming beam tests, the equilibration test isolates the implicit matter-coupling update by turning off transport and evolving only the local absorption/emission source in a homogeneous cell, for which the equilibrium and total energy are known.
We initialize the gas at a higher temperature than the radiation field, with \(\rho=1\), \(\Gamma=5/3\), \(T_{\rm gas}=2\), \(T_{\rm rad}=1\), \(a_{\rm rad}=1\), and pure absorption \(\kappa_a=1\).
The exact comparison solves the scalar energy-conserving ODE for \(T_{\rm gas}(t)\), subject to the conserved total energy:
\begin{equation}
  u_{\rm tot}
  =
  \frac{\rho T_{\rm gas}}{\Gamma-1}
  +
  a_{\rm rad}T_{\rm rad}^4.
\end{equation}

\begin{figure}[t]
\centering
\includegraphics[width=\linewidth]{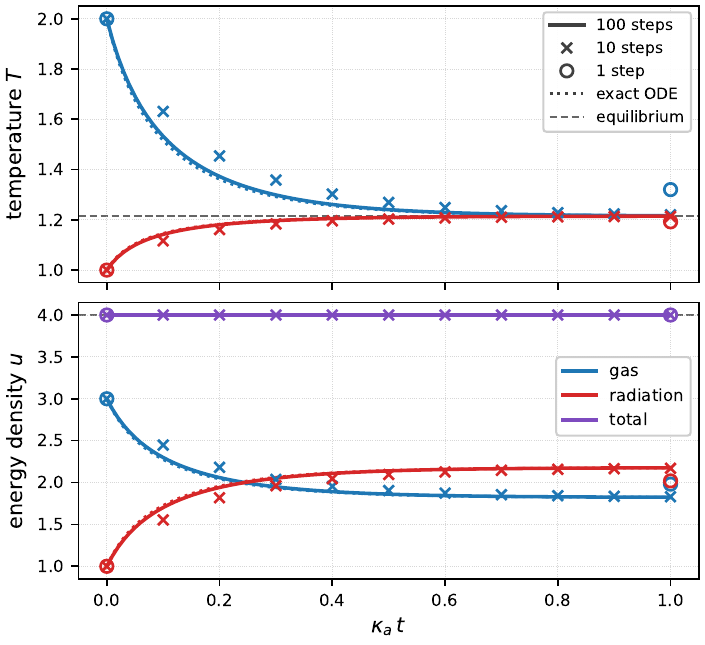}
\caption{Homogeneous gas--radiation equilibration compared with the exact energy-conserving ODE.
The panels show temperatures (top) and energy densities (bottom) for \(100\), \(10\), and \(1\) implicit source substeps.}
\label{fig:equilibration}
\end{figure}

For this test, a fixed source-evolution interval \(0\leq t\leq 1\) is divided into \(N_{\rm sub}=1\), \(10\), or \(100\) equal implicit source substeps.
As shown in Fig.~\ref{fig:equilibration}, the trajectory converges to the exact ODE as \(N_{\rm sub}\) is increased.
In the \(N_{\rm sub}=100\) run, the maximum temperature error is \(1.71\times10^{-2}\), the maximum energy error is \(2.57\times10^{-2}\), and the total energy is conserved to machine precision.

\subsection{Radiation-modified linear waves}

Radiation-modified linear waves test the coupled transport, fluid feedback, and matter update in the optically thick regime considered by \citet{White2023}, while the analytic eigenmode provides a direct convergence measurement.
We excite the radiation-fluid fast wave---the radiation-modified acoustic eigenmode of the coupled system---on a unit periodic domain containing one wavelength.
The background state is \((\rho,p_{\rm gas},E)=(1,0.2497687327,0.07493061980)\), with \(\Gamma=5/3\), \(a_{\rm rad}=19.25338273\), and \(\kappa_a=\kappa_s=10\).
We use a perturbation amplitude of \(10^{-6}\), WENO-Z reconstruction for the gas, piecewise-linear reconstruction for the radiation, RK2 time integration, and a CFL number of \(0.3\), and vary the resolution from \(N_{\rm cell}=32\) to \(4096\).

The final-time input is expressed in damping half-lives rather than in code time.
The adopted value of \(0.1\) therefore corresponds to \(t_{\rm final}=0.1\ln 2/|\operatorname{Im}\omega|=2.6466\).
For the complex eigenfrequency \(\omega=3.148815753-0.02619000639i\), this interval spans \(\operatorname{Re}(\omega)t_{\rm final}/(2\pi)=1.326\) wave periods and reduces the perturbation amplitude by the factor \(2^{-0.1}=0.933\).
At the end of each run, the problem generator evaluates the damped analytic eigenmode at the same time and phase as the numerical solution, so the comparison does not require an integer number of periods.

\begin{figure}[t]
\centering
\includegraphics[width=\linewidth]{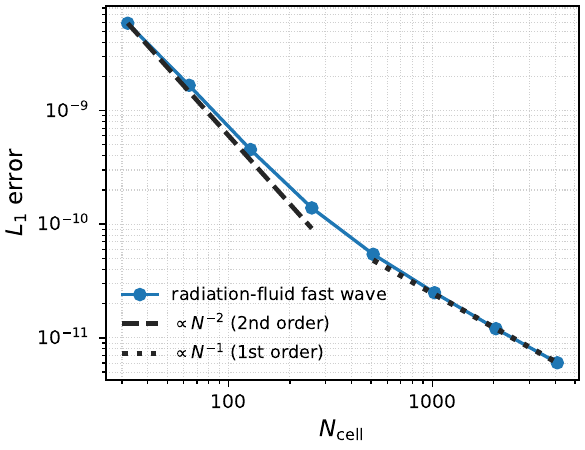}
\caption{Convergence of the radiation--fluid fast wave measured with the composite conserved-fluid \(L_1\) error.
The error is second order at low resolution and approaches first order once the operator-split radiation--matter coupling dominates.}
\label{fig:lwave_conv}
\end{figure}

For each conserved fluid component \(U_a\)---density, three momenta, and gas energy---we compute the volume-averaged absolute error \(\epsilon_{1,a}=V^{-1}\sum_i\Delta V_i|U_{a,i}^{\rm num}-U_{a,i}^{\rm exact}|\).
The y-axis in Figure~\ref{fig:lwave_conv} is the composite fluid error \(\epsilon_{\rm comp}=(\sum_a\epsilon_{1,a}^2)^{1/2}\), rather than the error of a single variable.
It does not include a separate norm of the radiation moments; instead, radiation transport and matter coupling affect the measured fluid error through the coupled evolution.
The HARM and Valencia radiation solvers give this diagnostic to six significant figures at every resolution.
At low resolution the error converges at close to second order, as expected from the spatial reconstruction and RK2 integrator.
As the grid is refined, the convergence rate approaches first order by \(N_{\rm cell}\simeq10^3\), when the first-order operator split between transport and the implicit radiation--matter source update dominates the remaining error.

\subsection{Analytic ADM tests}

The following two tests isolate geometric terms that are specific to the Valencia formulation. 
In each case, the metric is simple enough that the expected radiation evolution can be obtained analytically.

\begin{figure}[t]
\centering
\includegraphics[width=\linewidth]{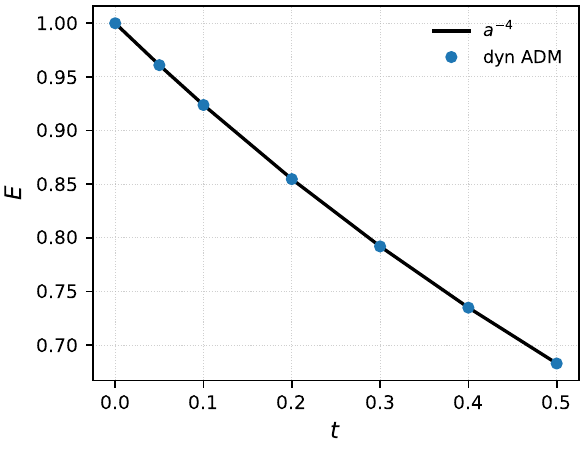}
\figcaption{Cosmological redshift on an expanding FLRW background compared with the analytic \(E\propto a^{-4}\) prediction.\label{fig:adm_flrw}}
\end{figure}

The first test uses a spatially flat FLRW cosmological spacetime \citep{Friedmann1922},
\begin{equation}
  \dd s^2
  =
  -\dd t^2
  +
  a^2(t)
  \left(
    \dd x^2+\dd y^2+\dd z^2
  \right),
\end{equation}
where we set the scale factor $a(t)=1+0.2t$. 
In ADM form, the lapse, shift, and spatial metric are
\begin{equation}
  \alpha=1,
  \qquad
  \beta^i=0,
  \qquad
  \gamma_{ij}=a^2(t)\delta_{ij}.
\end{equation}
For an initially isotropic, collisionless radiation field, cosmological redshift gives \(E\propto a^{-4}\), while the evolved densitized energy obeys \(\sqrt{\gamma}E\propto a^{-1}\).
We initialize \(E=1\) and use a unit periodic domain with an \(8\times4\times1\) mesh, \(N_{\rm ang}=42\), piecewise-linear reconstruction, RK2 time integration, and a CFL number of \(0.4\).
At \(t=0.5\), the relative error in \(E\) is \(1.19\times10^{-4}\), as shown in Figure~\ref{fig:adm_flrw}.

The second test isolates the response to a spatial lapse gradient on
a time-independent background,
\begin{equation}
  \begin{aligned}
    \dd s^2 =&
    -\alpha^2(x)\dd t^2
    +\dd x^2+\dd y^2+\dd z^2, \\
    &\alpha(x) = 1+0.1\sin(2\pi x).
  \end{aligned}
\end{equation}
At \(t=0\), the radiation field is uniform in space,
with \(E_0=1\), but anisotropic in angle. Its net flux points along
\(+x\) and has flux factor
\(f=|F^x_0|/E_0=0.7\).

\begin{figure}[t]
\centering
\includegraphics[width=\linewidth]{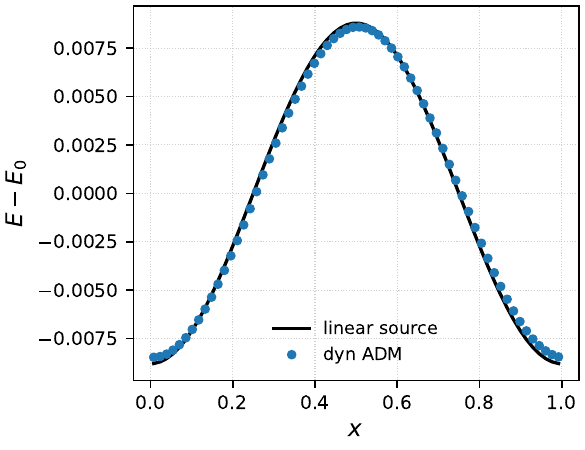}
\caption{Radiation-energy source from a static sinusoidal lapse gradient compared with the first-order analytic prediction.}
\label{fig:adm_lapse_gradient}
\end{figure}

Because \(\beta^i=0\), \(\gamma_{ij}=\delta_{ij}\), and \(K_{ij}=0\),
the radiation-energy equation becomes
\begin{equation}
  \partial_t E
  +\partial_x\!\left(\alpha F^x\right)
  =-F^x\partial_x\alpha .
\end{equation}
The right-hand side is the explicit lapse-gradient source. The
quantity plotted in Figure~\ref{fig:adm_lapse_gradient}, however, is
the total energy change \(E-E_0\), which also includes the divergence
of the transport flux. Since \(F^x_0\) is initially uniform, the two
contributions have equal magnitude, giving
\begin{equation}
  E(x,t)-E_0
    =-2F^x_0\partial_x\alpha\,t
      +\mathcal{O}(t^2).
\end{equation}

We use a unit periodic domain with a \(64\times4\times1\) mesh,
\(N_{\rm ang}=92\), piecewise-linear reconstruction, RK2 time
integration, and a CFL number of \(0.2\). In units with \(c=1\),
\(t=0.01\) is approximately \(1\%\) of the domain light-crossing
time. This short evolution time is chosen to test the first-order response; at \(t=0.01\), the
numerical profile of \(E-E_0\) has a correlation of \(0.999\) with
the first-order prediction shown in
Figure~\ref{fig:adm_lapse_gradient}.

\begin{figure*}[t]
\centering
\includegraphics[width=\textwidth]{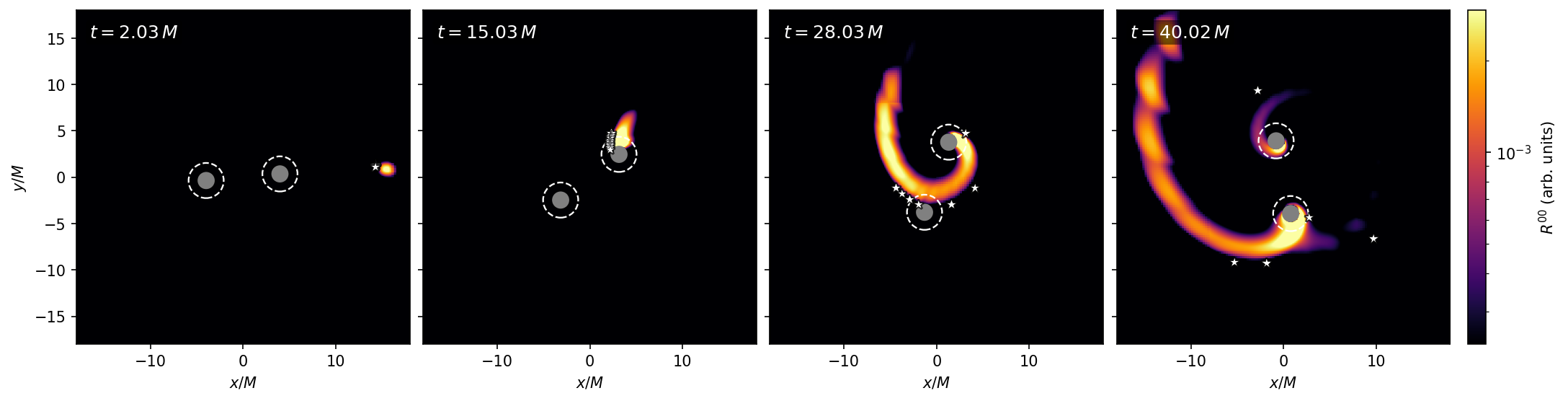}
\caption{Flash-beam transport on an orbiting equal-mass Kerr--Schild binary and a black-hole-tracking AMR grid at \(t/M=2.03,15.03,28.03,\) and \(40.02\).
Color shows the coordinate-frame component \(R^{00}\), with an independent logarithmic normalization in each panel to expose the pulse morphology.
White stars mark the instantaneous positions of the subset of ten null particles that are evolved in the same time-dependent geometry.
Gray disks mark the horizons, and dashed white circles mark the retrograde photon-ring radii, with the prograde ones close to the horizons.}
\label{fig:dynbbh_timeseries}
\end{figure*}

\subsection{Radiation beam on orbiting binary black holes}
\label{sec:dynbbh_metric}

To verify the Valencia radiation solver in the absence of any symmetry, we conduct a radiation ``flash'' beam demonstration that combines a time-dependent binary black hole metric with moving AMR and excision.
The binary spacetime follows the boosted-Kerr superposition introduced by \citet{Combi2021Metric,Ressler2024DynMetric}.
In the present implementation,
\begin{equation}
  g_{\mu\nu}
  =
  \eta_{\mu\nu}
  +
  h^{(1)}_{\mu\nu}
  +
  h^{(2)}_{\mu\nu},
\end{equation}
where each \(h^{(A)}_{\mu\nu}\) is an arbitrary-spin Cartesian Kerr--Schild perturbation centered on hole \(A\) and transformed by its instantaneous Lorentz boost along a prescribed circular orbit.
The \athenapp\ implementation evaluated the metric derivatives numerically \citep{Ressler2024DynMetric}.
Here forward-mode automatic differentiation is used within the metric provider to construct the four-metric derivatives and hence the ADM variables, including \(K_{ij}\).
The radiation module then forms the spatial derivatives required by angular transport by centered finite differences of the cached ADM fields.

The beam test uses two equal-mass holes with aligned dimensionless spins \(\chi_1=\chi_2=0.7\), pointing out of the page, at a coordinate separation of \(8M\), where \(M\) is the total mass of the binary.
It is run with \(N_{\rm ang}=162\) on a three-level AMR grid whose finest patches track the orbiting black holes, with the radiation field excised within the horizons.
A short exponential flash (\(e\)-folding time \(0.3M\)) is launched from a fixed point at coordinate radius \(16M\), aimed to graze the near hole, and its angular weights are re-projected onto the local angular grid every substep against the time-dependent metric.
After the flash shuts off the field free-streams.
The computational domain is the cube \([-24M,24M]^3\), covered by a \(96^3\) base grid and three total AMR levels, giving \(\Delta x=M/8\) on the finest level.
The run uses piecewise-linear reconstruction, RK2 time integration, and a CFL number of \(0.25\).
The flash has amplitude \(8\), spatial width \(0.375M\), an exponential decay time of \(0.3M\), and is active from \(t=M\) to \(2.5M\).

As an independent characteristic check, we initialize ten null particles at the fixed beam-launch point with directions uniformly spanning the full \(6^\circ\) opening angle, from \(-3^\circ\) to \(+3^\circ\) relative to the beam axis.
The particles are evolved by the null-geodesic pusher using the same time-dependent ADM geometry as the radiation solver.
They sample the angular deflection of the beam rather than its longitudinal extent: all ten particles are initialized simultaneously at one point, whereas the radiation source has finite spatial width and remains active from \(t=1\) to \(2.5M\).
Consequently, the particle positions are reference characteristics and are not expected to delineate the leading or trailing edge of the finite-duration radiation pulse at every time.

Figure~\ref{fig:dynbbh_timeseries} follows the pulse while the holes orbit each other.
The initially compact beam approaches and grazes one hole, is strongly deflected through the binary, and develops an extended curved wavefront by \(t\simeq40M\).
The instantaneous particle positions show the corresponding spread and strong-field bending of the sampled null characteristics.
Particles that enter a horizon are omitted, so fewer stars remain in the later panels.
The stars provide a geometric comparison with the transported radiation field; they are neither fitted to radiation features nor radiation contours.

\section{Demonstration: radiative accretion onto binary black holes}
\label{sec:application}

Black-hole binaries arise across a broad range of masses and astrophysical environments.
At the high-mass end, galaxy mergers are expected to produce bound massive black-hole binaries, whose demographics and accretion signatures encode the assembly history of galaxies and their central black holes \citep{Begelman1980,Bogdanovic2022Review}.
At the sub-parsec separations most relevant to their gravitational-wave-driven evolution, however, these systems are generally unresolved even when actively accreting.
Searches must therefore rely primarily on indirect signatures, including periodic or quasiperiodic photometric variability and relativistic Doppler modulation \citep{Graham2015Periodic,Charisi2016Periodic,DOrazio2015Doppler}.
Current and forthcoming time-domain surveys, including ZTF, Rubin/LSST, and Roman, will substantially expand these searches \citep{Bellm2019ZTF,Ivezic2019LSST,Haiman2023Roman}.
Despite this strong observational motivation, first-principles radiative models remain limited: existing angle-dependent radiation-MHD calculations have employed Newtonian gravity and treated either the circumbinary disk on scales outside the individual black holes or a single isolated minidisk at fixed binary separation \citep{Tiwari2025CBD,Chan2025Minidisk,2025arXiv251013955T}.
A global calculation that connects the circumbinary flow, the near-hole accretion structures, and the escaping radiation field in a dynamical relativistic binary spacetime has not previously been available.

As an end-to-end demonstration of the new transport method, we consider an exemplary accretion flow onto a stellar-mass binary black hole.
Gas-embedded stellar-mass binaries may arise within active galactic nucleus disks \citep[e.g.,][]{2012MNRAS.425..460M,Bartos2017AGN,Stone2017AGN}, or when the inner compact binary of a hierarchical triple is engulfed by the envelope of an evolved tertiary \citep[e.g.,][]{GlanzPerets2021}.
We do not attempt to model a particular formation channel and instead adopt generic initial conditions.
The stellar-mass scaling simplifies the problem, since the accreting gas should be fully ionized and therefore have opacity independent of its density and temperature,
while retaining the essential multi-regime transport problem: radiation is advected with the bulk flow and diffuses relative to the gas in optically thick regions, progressively decouples from the matter near the photosphere, and approaches free streaming in the optically thin polar funnels, all within a strongly time-dependent spacetime.

\subsection{Numerical setup and radiation initialization}
\label{sec:application_setup}

We consider an equal-mass binary with total mass \(M=40M_\odot\), coordinate separation \(d=25M\), and dimensionless spins \(\chi_1=\chi_2=0.7\) aligned with the orbital angular momentum.
The holes follow the prescribed circular trajectory used to construct the time-dependent superposed Kerr--Schild spacetime described in Section~\ref{sec:dynbbh_metric} \citep{Combi2021Metric,Ressler2024DynMetric,CombiRessler2026Metric}.
The orbital angular frequency and period are
\begin{equation}
  \Omega_{\rm bin}
  =
  d^{-3/2},
  \qquad
  P_{\rm bin}
  =
  \frac{2\pi}{\Omega_{\rm bin}}
  =
  785.4M.
  \label{eq:application_binary_period}
\end{equation}
For the adopted mass scale, \(P_{\rm bin}=0.155\,{\rm s}\), corresponding to an orbital frequency of \(6.46\,{\rm Hz}\) and a dominant gravitational-wave frequency of \(12.9\,{\rm Hz}\), near the nominal low-frequency edge of the Advanced LIGO sensitivity band \citep{Aasi2015AdvancedLIGO}.

\begin{figure*}[t]
\centering
\includegraphics[width=\textwidth]{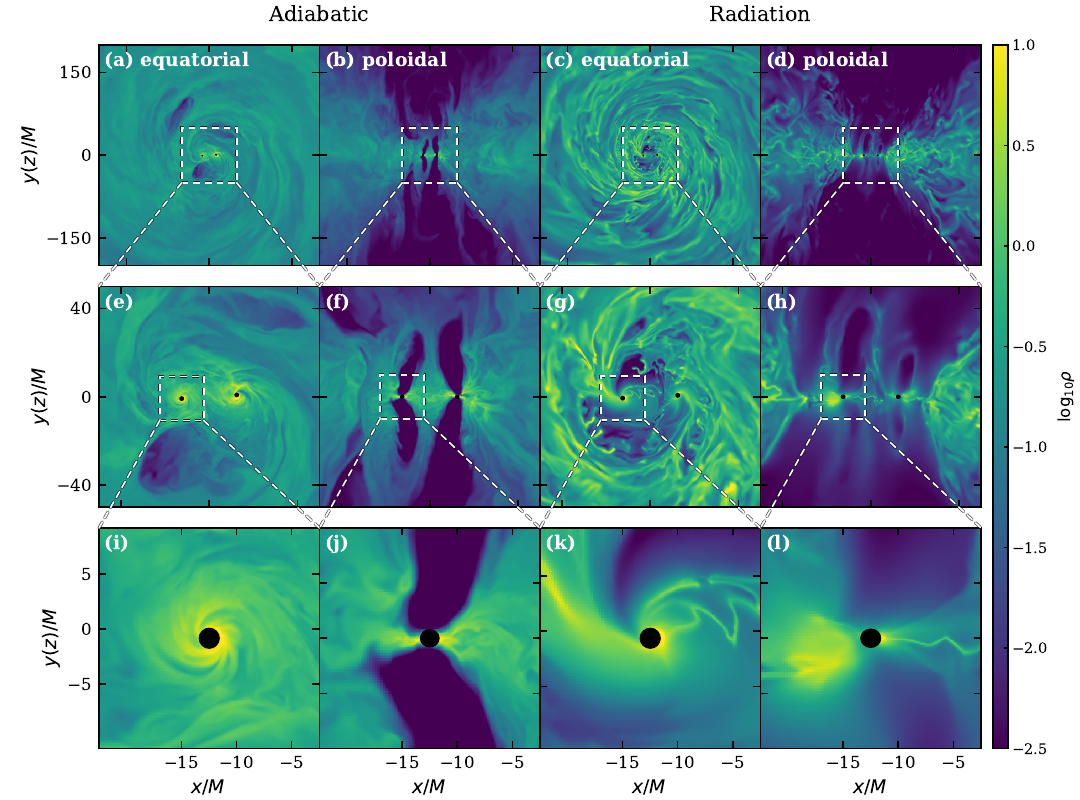}
\caption{Phase-matched equatorial and poloidal rest-mass density slices from the adiabatic GRMHD evolution (left columns) and its radiation-GRMHD continuation (right columns).
Rows span \(\pm200M\), \(\pm50M\), and \(\pm10M\), highlighting the circumbinary flow structure, the cavity, and accretion onto an individual hole respectively; black regions mark the excised horizon interiors.}
\label{fig:mad_rad_density}
\end{figure*}

The initial gas configuration is an extended Chakrabarti torus \citep{Chakrabarti1985} with inner edge at \(60M\), pressure maximum at \(240M\), and adiabatic index \(\Gamma=13/9\).
A net vertical magnetic field is initialized with minimum plasma beta \(\beta_{\min}=100\).
The torus is first evolved without radiation for approximately \(350\) binary orbits, until it develops a quasi-steady turbulent magnetically arrested accretion flow, or Binary--MAD (BMAD) \citep{Most2024,Most:2024onq,Ressler:2024tan,Wang2025yco}.
The computational domain extends to \(4096M\) and uses an outflow-only (``diode'') outer boundary condition.
A hierarchy of nested mesh-refinement levels surrounds the binary and tracks the two holes.
On the level containing the horizons, the grid spacing is
\begin{equation}
  \Delta x
  =
  \frac{M}{8}
  =
  0.125M,
\end{equation}
which resolves each \(\chi=0.7\) Cartesian Kerr--Schild horizon by approximately \(15\) cells across its equatorial diameter and \(14\) cells across its polar diameter.
For the GRMHD evolution, we use WENO-Z reconstruction, an HLLE Riemann solver, RK2 time integration with a CFL number of \(0.4\), and floors \(\rho_{\min}=10^{-10}\) and \(T_{\min}=10^{-8}\) in code units.

The conversion to physical units adopts the density normalization \(\rho_0=6.6\times10^{-6}\ {\rm g\,cm^{-3}}\).
For a \(40M_\odot\) system, this places the dense disk in the optically thick, electron-scattering-dominated regime relevant to luminous stellar-mass black-hole accretion.
The normalization is chosen to target a mildly super-Eddington flow.

Radiation is then activated from a checkpoint of the adiabatic evolution.
The discrete intensities are initialized to zero; an instantaneous locally-thermal-equilibrium radiation bath is not imposed because it would be inconsistent with the prior radiation-free evolution.
Absorption, emission, elastic scattering, transport, and radiation feedback are enabled immediately, allowing the radiation field to build up from the local emissivity of the existing flow.

Compton energy exchange is initially disabled after radiation activation, allowing the radiation energy to accumulate before the stiff Compton source term is introduced.
After one orbit, we then ramp up the Compton term over one orbital period using the smooth function:
\begin{equation}
  f_{\rm C}(s)
  =
  3s^2-2s^3,
  ~~
  s
  =
  \operatorname{min}
  \left[
    1,
    \operatorname{max}
    \left(
      0,
      \frac{t-t_0}{P_{\rm bin}}
    \right)
  \right],
  \label{eq:compton_ramp}
\end{equation}
where \(t_0\) denotes the start time of the Compton ramp.
The calculation is then continued through another five binary orbits.
This staged procedure separates the initial formation of the radiation field from the onset of efficient Compton exchange and avoids an impulsive thermal transient.

\subsection{Morphology of the radiative accretion flow}
\label{sec:application_morphology}

Figure~\ref{fig:mad_rad_density} compares phase-matched rest-mass density slices from the radiation-free BMAD evolution and its radiation-GRMHD continuation after approximately five binary orbits.
The rows successively enlarge the central \(200M\), \(50M\), and \(10M\), with the same density scale used throughout.
Since the radiative calculation continues from the adiabatic run, rather than from an independently evolved control, we restrict the discussion to the main morphological differences.

On the largest scale, the two flows remain broadly similar.
The circumbinary cavity survives after radiation is activated, and both runs retain an equatorially concentrated disk and low-density polar regions.
The radiation-GRMHD flow nevertheless contains more small-scale structure.
Its density field is broken into narrow filaments and compact overdensities distributed more evenly in azimuth, rather than the smoother spiral features seen in the adiabatic run.
The low-density channels also change shape, possibly reflecting a reorganization of the magnetic flux tubes and flux-eruption pattern \citep{Wang2025yco}.

This behavior is not well described as rapid local cooling.
The disk is optically thick and its photosphere lies well above the midplane, so much of the radiation is trapped and advected with the gas \citep{Zhang2026RadII}.
Radiation then contributes to the pressure and shifts the effective adiabatic index of the coupled gas--radiation fluid toward \(4/3\), below the gas value \(13/9\) used in the adiabatic evolution.
The softer equation of state makes the flow more compressible and provides a natural explanation for the increased density contrast and more irregular structure.

The difference becomes sharper in the \(50M\) panels.
The adiabatic cavity wall is relatively smooth, whereas the radiation-GRMHD run develops narrow, radially elongated fingers that penetrate the cavity.
Dense fingers separated by low-density channels are a characteristic signature of magnetic interchange in magnetically arrested circumbinary flows, where gas infiltrates inward as magnetic flux is displaced outward \citep{Most2024,Most:2024onq,Wang2025yco}.

The \(10M\) panels show the clearest qualitative change.
The adiabatic flow forms a coherent rotating structure around the hole, while the radiation-GRMHD flow is dominated by a highly anisotropic filament that reaches the horizon without forming a comparable minidisk.
Several effects could contribute.
The interchange streams may carry too little angular momentum to circularize, accumulated magnetic flux may intermittently disrupt the minidisk, or trapped radiation pressure and radiation forces may alter the stream shocks and angular-momentum transport.
Newtonian radiative-MHD calculations at different binary parameters have instead found thinner and denser minidisks \citep{Chan2025Minidisk}, so the suppression seen here should not be interpreted as a generic effect of radiation.
It may depend on the magnetic-flux state, optical depth, binary separation, or simply the orbital phase shown in Figure~\ref{fig:mad_rad_density}.
We defer detailed analysis to future work.

\begin{center}
\begin{minipage}{\linewidth}
\centering
\includegraphics[width=\linewidth]{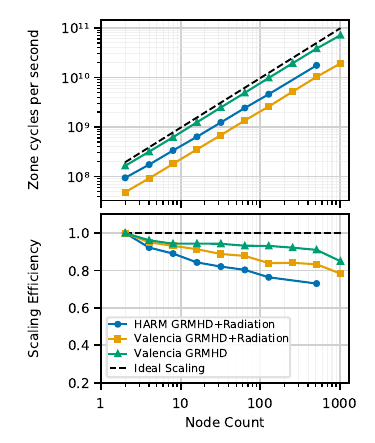}
\figcaption{Weak scaling on a fixed-spacetime torus benchmark for the HARM radiation-MHD, Valencia radiation-GRMHD, and radiation-free Valencia GRMHD configurations.
The Valencia radiation calculation reconstructs the ADM geometry and tetrad at every RK substep.
The panels show aggregate throughput (top) and efficiency relative to the two-node result of each configuration (bottom).\label{fig:weak_scaling}}
\end{minipage}
\end{center}

\section{Weak scaling}
\label{sec:weak_scaling}

We measure the weak scaling of the radiation solver on Aurora at the Argonne Leadership Computing Facility.
Each Aurora node contains two Intel Xeon CPU Max Series processors and six Intel Data Center GPU Max Series 1550 GPUs, with two compute tiles per GPU and \(768\,\mathrm{GB}\) of total GPU high-bandwidth memory.
The nodes are connected by the HPE Slingshot 11 network.
We increase the node count from 2 to 1024 while keeping the mesh and angular workload per node fixed.

The benchmark uses the same stationary torus spacetime for all three configurations.
For the Valencia radiation calculation, however, the ADM quantities and Cholesky tetrad are reconstructed at every RK substep, following the same path used when the metric is time-dependent.
The timing therefore includes the geometry-refresh cost of dynamical-spacetime radiation transport, but not the additional cost of evolving the spacetime with Z4c.
This distinction is important when comparing the HARM and Valencia results: the HARM geometry is precomputed, whereas the Valencia geometry is rebuilt throughout the evolution.

Figure~\ref{fig:weak_scaling} compares the aggregate zone-cycles per second and the efficiency normalized to the two-node result of each configuration.
All three configurations maintain good weak scaling over the measured range.
At the largest node count available for each series, the efficiency is \(0.73\) for HARM radiation-MHD on 512 nodes, \(0.78\) for Valencia radiation-GRMHD on 1024 nodes, and \(0.85\) for radiation-free Valencia GRMHD on 1024 nodes.
The HARM radiation solver remains faster at fixed node count, although the throughput ratio decreases from \(1.95\) on two nodes to \(1.71\) on 512 nodes.
The somewhat better scaling of the Valencia radiation solver is consistent with its larger amount of local work from reconstructing the geometry and tetrad at every substep.

Radiation lowers the Valencia throughput by factors of \(3.42\) on two nodes and \(3.70\) on 1024 nodes relative to the radiation-free Valencia baseline.
These measurements compare complete simulation configurations rather than individual radiation kernels.

\section{Summary and outlook}
\label{sec:summary}

We have extended the finite-solid-angle general relativistic radiation transport method of \citet{White2023} to time-dependent spacetimes represented in ADM form.
The new Valencia radiation solver evolves a densitized Eulerian intensity, constructs its local tetrad algebraically from the spatial metric, and computes angular geodesic advection from the ADM Hamiltonian.
Its geometric terms require only the instantaneous lapse, shift, spatial metric, extrinsic curvature, and their spatial derivatives, allowing the transport scheme to couple directly to the ADM/Z4c infrastructure in \athenak.

The tests verify spatial and angular transport, strong-field light bending, radiation--matter coupling, and the geometric terms introduced by an evolving spacetime.
On stationary backgrounds, the Valencia and HARM solvers converge to the same physical solution after accounting for their different normalization of conservative variables and tetrad gauges.
The binary-black-hole beam test demonstrates transport on a time-dependent, black-hole-tracking AMR grid, while the radiative MAD calculation shows that the full method can be sustained in a turbulent, optically thick binary accretion flow with moving excision regions and stiff matter coupling.
The implementation retains good weak scaling to the largest configurations tested, with \(1024\) GPU nodes on Aurora hosted at the Argonne Leadership Computing Facility.

The present implementation is gray (frequency-integrated), and the Lie splitting between transport and matter coupling limits the asymptotic temporal accuracy of coupled problems to first order.
Natural extensions include multifrequency (multigroup) transport, higher-order source integration, broader opacity models, and extension to neutrino opacities and coupling.
This work provides a performance-portable implementation of deterministic Boltzmann transport for dynamical relativistic systems.

\begin{acknowledgments}

We thank Luciano Combi, Shane Davis, Jacob Fields, Yan-Fei Jiang, Elias Most, Patrick Mullen, Frans Pretorius, Hai-Yang Wang, and Chris White for insightful discussions.
H.Z. acknowledges the use of \textbf{Codex} by OpenAI to accelerate numerical implementation and manuscript preparation.
All scientific analyses, numerical results, and conclusions were produced by the authors, who take full responsibility for the contents of this work.
We further thank our INCITE coordinator Kyle Felker and Michael Buehlmann for their extensive assistance with running on the Aurora Exascale Supercomputer hosted at ALCF.
An award of computer time was provided by the U.S. Department of Energy's (DOE) Innovative and Novel Computational Impact on Theory and Experiment (INCITE) Program, under projects \textbf{RadBlackHoleAcc} and \textbf{CompactBinaryMerger}, as well as the Director's Discretionary project \textbf{BBHGRMHD}.
This research used resources from the Argonne Leadership Computing Facility, a U.S. DOE Office of Science user facility at Argonne National Laboratory, which is supported by the Office of Science of the U.S. DOE under Contract No. DE-AC02-06CH11357.
Early test simulations for this paper were also performed on computational resources managed and supported by Princeton Research Computing, a consortium of groups including the Princeton Institute for Computational Science and Engineering (PICSciE) and Research Computing at Princeton University.

A.J.D. was supported by NASA through the NASA Hubble Fellowship grant No. HST-HF2-51553.001, awarded by the Space Telescope Science Institute, which is operated by the Association of Universities for Research in Astronomy, Inc., for NASA, under contract NAS5-26555.
E.M.G. acknowledges support from the National Science Foundation under Grant No. PHY‐2407681. D.R. acknowledges support from NASA through Grant No. 80NSSC25K7213, from the Department of Energy, Office of Science, Division of Nuclear Physics, under Award Number DE-SC0024388, and from the National Science Foundation under Grants No. PHY-2512802 and PHY‐2621752.  J.S. acknowledges support from NASA Grant No. 80NSSC25K7213 through a subcontract to PSU.
\end{acknowledgments}

\bibliographystyle{aasjournalv7.1}
\bibliography{references}

\end{CJK*}
\end{document}